\documentclass[,final]{aipproc}
\layoutstyle{6x9}

\def\be{\begin{equation}}
\def\ee{\end{equation}}
\def\ba{\begin{eqnarray}}
\def\ea{\end{eqnarray}}

\def\negenspace{\kern-1.1em}

\begin{document}
%
\title{Einsteinian gravity from a topological action}
\classification{04.20.Fy, 04.60-m, 04.50.+h, 11.15-q}
\keywords{Topological gravity, double duality, BRST quantization,
Einsteinian background}
\author{Eckehard W. Mielke\thanks{E-mail: ekke@xanum.uam.mx}}{
address={Departamento de F\'{\i}sica,\\
Universidad Aut\'onoma Metropolitana--Iztapalapa,\\
Apartado Postal 55-534, C.P. 09340, M\'exico, D.F., MEXICO}}

\maketitle
\begin{abstract}
The  curvature-squared model of gravity, in the affine form proposed
by Weyl and Yang, is deduced from a topological action in 4D. More
specifically, we start  from the Pontrjagin (or Euler) invariant.
Using the BRST antifield formalism with a double duality {\em gauge
fixing}, we obtain a consistent quantization in spaces of double
dual curvature as classical instanton type background.

However,  exact vacuum solutions with double duality properties
exhibit a `vacuum degeneracy'. By modifying the duality via a {\em
scale breaking} term, we demonstrate that only Einstein's equations
with an induced {\em cosmological constant} emerge for the {\em
topology} of the macroscopic background. This may have repercussions
on the problem of `dark energy' as well as `dark matter' modeled by
a torsion induced quintaxion.
\end{abstract}


\section{Introduction}
On the macroscopical level, Einstein's general relativity has passed
every test ``with flying colors", see Will \cite {Wi05} for a recent
review. However, Einstein's theory sofar has resisted any attempt
of quantization, e.g. it is known to be perturbatively
non-renormalizable, partially due to its dimensional  coupling
constant\footnote{It has to be kept in mind that Newton's
gravitational constant $G$ is one of the less precise known
constants of physics. In order to improve this situation, there are
plans  \cite{AB01} to measure the gravitational attraction of two
bodies in a spaceship (Project SEE), where the larger body will
function as a shepherd for the movement of the test mass, similarly
as in the rings of Saturn. Amongst others, an accuracy of $3.3\times
10^{-7}$ in the determination of $G$ should be feasible.}. String
theory or brane scenarios with extra dimensions have been proposed
as a rescue, some of which are implying, however,  deviations of
standard gravity in the sub-millimeter range. Recent torsion balance
experiments \cite{KC07} have probed the inverse square law and found
no deviation even below the hypothetical dark energy (DE) scale of
$\lambda_{\rm DE}=(\hbar c/\rho_{\rm DE})^{1/4}\simeq 85\; \mu$m.
Thus this  `window' of possibly new gravitational  physics seems
also to be closing.

In 1974  Yang proposed an affine gauge theory gravity \cite{Ya74}
which, due to its scale invariance, can be regarded as a rather
promising fundamental theory of (quantum) gravity in the high-energy
limit \cite{HMMN89}, without invoking extra dimensions or
supersymmetry, cf. Ref. \cite{KS86}. An additional duality
constraint on the curvature could be extremely important for the
path integral approach to quantum gravity. Then
 {\em instanton} type configurations \cite{GH78} near the
classical ones, i.e. Einstein spaces, are more  probable then the
`spurious' Thomson spaces, as one would expect naively. {}For the
modified duality with a breaking of  scale invariance, the
transition amplitude  peaks at classical Einstein spaces {\em only}.
Alternatively, in a four-dimensional Yang-Mills theory  gauging the
de Sitter group \cite{MM77,Pa83},  scale invariance gets
spontaneously broken by a pseudo-Goldstone type `radius vector'
\cite{TM00}, odd under CP, in order to  recover the Hilbert-Einstein
action plus the Euler term.

{}From the work of Stelle \cite{St77} we know that the curvature
squared gravity in Riemannian spacetime is perturbatively {\em
renormalizable} but unfortunately plagued with {\em physical} ghost,
i.e. negative residues in the graviton propagator \cite{LN90}. This
finding has diminished the initial interest \cite{Fa76,Fa77} in such
models.

Much more promising and elegant is to start from a purely
topological classical action, proportional to the gravitational
Pointrjagin (or Euler) invariant and then quantize this model by
nilpotent {\em Becchi-Rouet-Stora-Tyutin} (BRST) transformations
generated by $s$. Such a topological action is not only completely
metric-free, but moreover conformally invariant \cite{WiG88,LP88}
and can provide a consistent {\em Topological Quantum Field Theory}
(TQFT) as shown by Witten \cite{Wi88,BB91}. Lateron, it was realized
by Baulieu and Singer \cite{BS88} that for Yang-Mills theory
Witten's action is a gauge-fixed version of the classical
topological action through the standard BRST quantization procedure.
Then the total Lagrangian consists of a d-exact part, as exemplified
by the Pontrjagin invariant of the structure group, as well as an
$s$-exact piece accounting for the chosen gauge fixing. Modulo an
exact form, the full  Lagrangian turns out to be BRST-invariant.

In the case of gravity one can obtain a rather realistic
gravitational `background' dynamics, if we complete the topological
action by  {\em gauge constraints}, enforcing the Lorentz condition
on the linear connection and  {\em double duality} on the curvature.
Then the resulting model depends on the metric of spacetime only via
the $s$-exact term. This conforms with the general expectation that
it is ultimately the process of quantization which necessarily
induces a physical scale into a primordial topological model.
According to Faddeev \cite{Fa96}, quantization amounts to a {\em
stable deformation} of the classical Poisson algebra with a
dimensional transmutation due to the physical dimension $[\hbar]
=[p][q]$ of Planck's constant $\hbar$.

 In this paper, we investigated the BRST
quantization of topological gravity, cf. Ref. \cite{WiG88},
following essentially Baulieu and Singer \cite{BS88}, cf. also Refs.
\cite{Ba85,CB92,BT02}. Moreover, its classical limit,
 corresponding to  the most probable, extremal `trajectories'
in the Feynman path integral, is analyzed. In the case of gravity,
these are classical configurations with  self- or anti-self dual
curvature. In order to lift this `vacuum degeneracy', a modified
double duality constraint is considered which explicitly breaks
scale invariance. For torsionless configurations, we can demonstrate
that only Einstein's GR, consistently coupled to the symmetrized
energy-momentum current of matter fields, surface as low-energy
(long range) effective theory, satisfying all macroscopic tests. The
paper focusses on the main physical features of the model, deferring
details of the formalism  to three Appendices.

\section{Weyl-Yang theory of gravity}
 In 1974
Yang \cite{Ya74} tentatively considered the possible replacement of
Einstein's general relativity (GR) by an {\em affine} gauge theory
with a Yang-Mills type action. In fact, curvature-squared
Lagrangians had been considered before, first in 1919 by Weyl
\cite{We19} with the emphasis on scale invariance, and then later by
Stephenson \cite{St58}, Higgs \cite{Hi59}, as well as Kilmister and
Newman \cite{KN61}.

When written in differential forms,
the Stephenson-Kilmister-Yang (SKY) Lagrangian is given by
\begin{equation}
L_{\rm SKY}=-\frac{1}{2}R^{\alpha\beta}\wedge\,^*R_{\alpha\beta}\, .
\label{SKY}
\end{equation}
Our notation follows closely Cartan's exterior calculus, cf. Ref.
\cite{HMMN95} for details.  Then Yang's vacuum equations
\begin{equation}
D\,^*R_{\alpha\beta}=0 \label{Yangeq}
\end{equation}
together with \be E_\alpha=\frac{1}{2}\left( e_\alpha\rfloor
R^{\mu\nu}\wedge \,^*R_{\mu\nu} - R^{\mu\nu} \wedge e_\alpha\rfloor
\,^*R_{\mu\nu}\right)=0 \label{genergy}\ee follow from varying the
Lagrangian with respect to the  connection one-form
$\Gamma^{\alpha\beta}:= \Gamma_i{}^{\alpha\beta} dx^i$ as well as
with respect to the coframe $\vartheta^{\alpha} =
e_{j}{}^{\alpha}\,dx^{j}$.

The short-time initial value problem of Eq. (\ref{Yangeq}) is
well-posed \cite{GN98}. Moreover,  it does not depend on any length
scale, i.e. it is scale invariant as envisioned by Weyl \cite{We29}.
In this aspect it would be a good starting point for quantization,
if one could avoid the physical ghost in the propagator of the
gravitons. Consequently, a related purely topological action, in
which the Hodge dual $^*$  in (\ref{SKY}) is dismissed, could be
more promising.

\section{Topological action}
Adopting ideas of Witten \cite{Wi88} for a topological Yang-Mills
theory (TYM) and replacing  the internal $SU(N)$ group by the linear
group $SL(4,R)$ of the tangent space embracing the Lorentz group
$SO(1,3)$ (or $SO(4)$ in Euclidean space with signature ${\rm sig}=0$) as
subgroup, we start from the gravitational Pontrjagin four-form \ba
 L_{\rm Pontr}&:=& dC_{\rm RR}={\rm Tr}\{\Omega\wedge\Omega\}\nonumber\\
 &=&
 {1\over 2}R_{\alpha}{}^{\beta}\wedge R_\beta{}^{\alpha}
=\frac{1}{2} d \big( \Gamma_{\alpha}{}^{\beta}\wedge
R_\beta{}^{\alpha} - {1\over 3} \Gamma_\alpha{}^\beta\wedge
\Gamma_\beta{}^\gamma\wedge \Gamma_\gamma{}^\alpha\big)\, ,
\label{Pontr} \ea which locally is a d-exact form violating parity
$P$. Our 'Clifform' notation is summarized in Appendix A following
Ref. \cite{Mie01}. This Lagrangian is completely metric-free and as
well invariant under the topological BRST transformations $s$ modulo
an exact form, i.e.
\begin{equation}
 sL_{\rm Pontr}=2d {\rm Tr}\{\Psi\wedge\Omega\}=
 d(\Psi_{\alpha}{}^{\beta}\wedge R_\beta{}^{\alpha})\,.
 \label{sinvar}
\end{equation}
Here $\Psi$ the one-form of the {\em topological ghost} with values
in the Lie algebra of the linear group and the topological BRST
transformations $s$ are specified in Appendix B. A similar result
would hold for the CP-invariant Euler term (\ref{Euler}), i.e.
$sL_{\rm Euler}=(-1)^{{\rm sig}}d\left(\Psi^{\alpha\beta}\wedge
R_{\alpha\beta}^{(\star)}\right)$. However, the latter is only
partial metric-free, since it involves the signature ${\rm sig}$ of
the metric implicitly in the definition of the Lie dual, cf. Ref.
\cite{BT90}.

This purely topological action can be amended by any $s$-exact form,
provided $s$ is a nilpotent BRST transformation, i.e. one with
$s^2=0$ without affecting the nice properties of the topological
action. Following Baulieu and Singer \cite{BS88}, we may choose the
Lorentz type conditions $d\,^* \Gamma=0$, $D\,^* \Psi=0$ on the
connection and the topological ghost, respectively, as well as  the
self-or anti-selfduality $\,^{\pm}\!\Omega ^{(\pm)}=0$ of the
curvature as {\em gauge constraints} consistently implemented via
the Faddeev-Popov type Lagrangian
\begin{eqnarray}
L_{\rm FP} &:=&- s {\rm Tr}\left\{ \overline{\chi} \wedge
\,^{*\pm}\!\Omega ^{(\pm)} + \overline{\Phi} D\,^* \Psi +
 \frac{1}{2}\rho \overline{\chi} \wedge \,^*b +\overline{c} d\,^* \Gamma
+ \frac{1}{2} \overline{c} \wedge \,^*B \right\} \nonumber \\ &=&
\frac{1}{2} s\left[ \overline{\chi}^{\alpha\beta} \wedge
\,^{*\pm}\!R_{\alpha\beta}^{(\pm)} + \overline{\Phi}_{\alpha\beta}
D\ ^* \Psi^{\alpha\beta} + \frac{1}{2}\rho
\overline{\chi}^{\alpha\beta} \wedge\,^* b_{\alpha\beta}
+\widetilde\Psi \right]\, , \label{FP}
\end{eqnarray}
where the trace is over the generators of the Lie algebra. Such a
constraint, at the same time, constricts the linear group $SL(4,R)$
to the Lorentz (or 4D orthogonal) group as a subgroup. For
comparison, also its component form is given. In order to end up
with a four-form Lagrangian, the auxiliary fields $\overline{\chi}$
and $b$ have to be selfdual two-forms. In this framework, the
so-called fermionic constraint is
 $\widetilde\Psi:= \overline{c}_{\alpha\beta} d\ ^*
\Gamma^{\alpha\beta} +\frac{1}{2}\overline{c}\wedge \, ^*B $.

In the following we are going to demonstrate that the full
topological gravity Lagrangian \be L_{\rm TG} =dC_{\rm RR}+ L_{\rm
FP}\ee is BRST invariant and classically equivalent to the self- or
anti-selfdual version of SKY gravity as given by Eq. (\ref{SKY}).

\section{Effective selfdual SKY gravity}
Performing the BRST transformation $s$ in the gauge-fixing
Lagrangian (\ref{FP}), we obtain after a long but straightforward
calculation using the rules of Appendix B the result

\ba L_{\rm FP} =& -& {\rm Tr} \Bigl\{ b \wedge
\,^{*\pm}\Omega^{(\pm)} + \overline{\chi} \wedge \,^*D\Psi -
\overline{\chi} \wedge\,^*
\left[c,\,^\pm\Omega^{(\pm)}\right]\nonumber\\
&+& \overline{\eta} D\,^*\Psi + \overline{\Phi}\,D\,^*D\Psi
+\overline{\Phi}\,[\Psi,^*\Psi] +B\,d\,^*\Gamma  \nonumber\\ &-&
\overline{c}\, d\,^*\Psi + \overline{c}\, d\,^*\!(Dc)  + \frac{1}{2}
\rho\, b \wedge \,^*b + \frac{1}{2} B \wedge {}^*\!B \Bigr\}\, . \ea
The variation with respect to the auxiliary field $B$ yields the
Lorentz type condition $d\,^*\Gamma =\,^*B$ on the connection.
Moreover, for vanishing real {\em gauge parameter} $\rho=0$, the
equation of motion for the two-form $b$ enforces the self- or
anti-self double duality condition
\begin{equation}
\,^{\pm}\!\Omega ^{(\pm)}:= \Omega \pm \,^*\Omega^{(\star)} =0
\label{DDm}
\end{equation}
on the curvature two-form $\Omega$, where we distinguish between the
Hodge dual $^*$ and the Lie dual $^{(\star)}$ in a space(-time) of
signature ${\rm sig}$.

In the case of the choice $\rho=1$ of the real gauge parameter, the
two-form $b$ is present in two terms, but then can be eliminated by
a Gaussian integration in Euclidean space such that, up to gauge
fixing terms, the SKY Lagrangian remains supplemented by the
topological Euler term (\ref{Euler}) as a boundary term, i.e.,
 \begin{eqnarray}
L^{(\star)} _{\rm SKY}&=& \frac{1}{2} {\rm Tr}\left(\,^{\pm}\!\Omega
^{(\pm)}\wedge\,^{*\pm}\!\Omega ^{(\pm)} \right)\nonumber\\
&=&-\frac{1}{2}R^{\alpha\beta} \wedge\,^*R_{\alpha\beta} \mp
\frac{(-1)^{\rm sig}}{2}R^{\alpha\beta}\wedge R_{\alpha\beta}^{(\star)} \nonumber\\
&=& -\frac{1}{4} \left(R_{\alpha\beta} \pm
\,^*R_{\alpha\beta}^{(\star)}\right)\wedge
\,^*\left(R^{\alpha\beta}\pm \,^*R^{\alpha\beta(\star)} \right)
 \,. \label{SKYmod}
\end{eqnarray}
The following argument is similarly to the path-integral approach to
gravity, where the transition amplitude $\int {\cal D}\Gamma
\exp\bigl[-\int V^{(\star)}_{\rm SKY}d^4x/\hbar\bigr]$
 is evaluated in an imaginary `spacetime' with Euclidean signature,
 cf. Ref. \cite{Mi84b}:
  It is obvious from
the equivalent binomial form of the effective SKY Lagrangian that
anti-selfdual solutions \cite{Mi81,BD81,MR05} as well as
 the selfdual spaces
\begin{equation}
  R_{\alpha\beta}= \mp \,^*R_{\alpha\beta}^{(\star)} \label{DDp}
\end{equation}
 are extrema, but the minimum contributes the most to the transition amplitude. (For Euclidean signature ${\rm sig}=1$, their solutions
 correspond to gravitational instantons, cf. Ref. \cite{EGH80}.) Both constraints
 satisfy Yang's
equation (\ref{Yangeq}) due to the Bianchi identity (\ref{Bia}) for
the Lie dual of the curvature as well as the condition
(\ref{genergy}) of vanishing gauge field energy, typical for
instantons. According to (\ref{DDual}) the first case corresponds to
Einstein spaces, {\em annihilating} the corresponding partially
topological action, whereas the selfdual case induces Thompson
spaces \cite{Th75}. This `vacuum degeneracy' was known already to
Fairchild \cite{Fa76,Fa77} and discussed in more detail in Ref.
\cite{Va02}, where the anti-selfdual SKY model has been coined
``Yang-Mielke" theory of gravity.

Concentrating on  topological terms such as those of
Pontrjagin (\ref{NiehYan}) and   Euler  (\ref{Euler}), related
 self--dual modifications have also been advocated as
{\em topological 4D selfdual gravity} by Nakamichi et al.
\cite{NSO91}. There, self- or anti-selfdual solutions are `living'
on Einstein spaces, as well. Moreover, deformations of conformal
gravitational instantons can be classified topologically, cf.  Ref.
\cite{PT93}. The addition of the Pontrjagin term with respect to the
Riemannian curvature $R^{\{\}}_{\alpha\beta}$ and the {\em axial
torsion} one-form ${\cal A}:= \,^*(\vartheta_\alpha\wedge T^\alpha)$
is rather well motivated by the  {\em axial anomaly} \be\langle
dj_5\rangle=  2i\,^*m \langle \overline{\psi}\gamma_5\psi \rangle
-\frac{1}{48\pi^2} \left(R^{\{\}}_{\alpha\beta}\wedge
R^{\{\}\alpha\beta}  + \frac{1}{2} d{\cal A}\!\wedge\!d{\cal
A}\right)\label{chianol}\ee in the coupling to Dirac fields $\psi$,
cf. Refs. \cite{KM01,MiA06} and the literature therein.

\section{Modified double dual gauge fixing}
In order to lift the vacuum degeneracy of selfdual SKY gravity, we
may impose instead a modified gauge constraint via \be L_{\rm FP}
(R) = s {\rm Tr} \Bigl\{\overline{\chi} \wedge \left(H -\theta_{\rm
L}^\star\, \Omega^{(\star)} - {{\theta_{\rm T}^*}\over{2\ell^2}}\,
\sigma\right)\Bigr\}\, , \ee
 involving
the curvature excitation  $H:=
\frac{i}{4}H_{\alpha\beta}\sigma^{\alpha\beta}$ associated with a
general {\em quadratic} curvature Lagrangian $L_{\rm QPG}$ in
Poincar\'e gauge framework, where \be H_{\alpha\beta}
 := -\partial L_{\rm QPG}/\partial R^{\alpha\beta} = -\,^*\left(
  \sum_{N=1}^{6}b_{(N)}\,^{(N)}R_{\alpha\beta}\right) \label{QR}
  \ee
  can be expanded into the irreducible curvature pieces.
 The propagating modes and  particle content of such a model are known from the work
of Sezgin and van Nieuwenhuizen \cite{Peter}, cf. Ref. \cite{Ku86}.
In addition, the Hilbert-de Donder gauge condition $d^*\gamma =0$
and further ghost constraints need to be added to comply with the
BRST algebra of the coframe $\gamma:=\vartheta^\alpha \gamma_\alpha$
indicated in Appendix C.

Then there arises  the modified {\em double duality ansatz}
\begin{equation} H_{\alpha\beta}(**) =
 \theta_{\rm L}^\star\, R^{(\star)}_{\alpha\beta}  +
{{\theta_{\rm T}^*}\over{2\ell^2}}\; \eta_{\alpha\beta}
\label{dodual}
\end{equation}
for the rotational field momenta \cite{Mie84,Mie91,Zhytnikov} as a
gauge constraint. Here $ \theta_{\rm T}^*$, and $\theta_{\rm
L}^\star$  are dimensionless constants  related to the individual
coupling constants in the $\theta$--type boundary terms
(\ref{RPontr}) and (\ref{Euler}). (The instanton solutions of Yang's
theory of gravity, classified \cite{Mi81} already 1981, are a
special case of the ansatz (\ref{dodual}) for the choice
$\theta_{\rm T}^*=0$ and $\theta_{\rm L}^\star=\mp 1$.) Interesting
enough, it can be regarded as {\em field redefinition} (FR) of the
linear connection $\Gamma$ such that (\ref{dodual}) is induced, see
Ref. \cite{Mi06} for details\footnote{In a rather ad hoc fashion,
such a FR was applied in Ref. \cite{OH96} to Euler and Pontrjagin
type terms. However, such
 deformations change the latter four-forms from being
anymore d-exact terms, thus prevailing a topological
interpretation.}.

\section{Einstein equation with induced cosmological constant
and axidilaton coupling}

In accordance with the classical field equations of affine gauge
theory, in the torsionless case we are then left with Eq. (5.8.29)
of Ref. \cite{HMMN95}, i.e.
 \begin{equation}- E_\alpha = \frac{\theta_{\rm T}^*}{2} R^{\{\}\beta\gamma}\wedge
\eta_{\alpha\beta\gamma} -\theta_{\rm T}^*
\Lambda_\theta\,\eta_{\alpha} = \ell^{2}\,\sigma_{\alpha}\, ,
\label{Eintheta}
\end{equation}
where $G:= \frac{1}{2}R^{\beta\gamma}\wedge
\eta_{\alpha\beta\gamma}\gamma^\alpha$ is the usual Einstein-Cartan
(EC) current three-form which is dual to the usual Einstein tensor
$G_{ij}:= Ric^{\{\}}_{ij}-\frac{1}{2}g_{ij}$. In the case of the
real gauge parameter $\rho=1$, the two-form $b$  can again be
eliminated by a Gaussian integration such that, up to gauge fixing
terms, a generalization of the dual SKY Lagrangian similar to the quadratic expression
(9.8) of Ref. \cite{Mi84b} remains.

In both cases, our gauge constraints induce the classical Einstein
equations
 \begin{equation}G_\alpha
- \Lambda_\theta\,\eta_{\alpha} =\frac{ \ell^{2}}{\theta_{\rm
T}^*}\,\sigma_{\alpha} \label{Ein}
\end{equation}
for the Riemannian background with the {\em symmetric}
Belinfante-Rosenfeld three-form $\sigma_{\alpha}:=\Sigma_\alpha
-D^{\{\}}\mu_\alpha$ as source and an effective gravitational
coupling constant $\kappa_{\rm eff}= \ell^{2}/\theta_{\rm T}^*$.

Thus by  modifying the gauge constraint,
the double duality relation (\ref{dodual}) surfaces which eliminates
the `vacuum ambiguity' for the exact solutions of SKY gravity, and
 only Einstein spaces remain as classical `background'.
 Due to the explicit appearance of a
length scale $\langle \varphi \rangle \propto 1/\ell$ in the ansatz
(\ref{dodual}), it is suggestive to associate this  with a
(sponteneous)  {\em symmetry breaking} of the  scale or Weyl
invariance of the original Lagrangian $L_{\rm QPG}$ related to
(\ref{QR}), for instance in a model \cite{HMMN89} dynamically
coupled to  a dilaton field $\varphi$, cf. Ref. \cite{DT02}. In a
Riemann-Cartan-Weyl spacetime, generalizations of  Einstein's
equations with axial torsion  as exemplified in Eq. (\ref{Eindec})
and a Weyl covector coupling can arise. When they are induced by an
axion $a$ and dilation $\varphi$ as potentials, a cancelation of the
axial torsion part in the chiral anomaly (\ref{chianol}) can be
achieved. Similarly as in strings, both may even combine into a
single complex scalar, the {\em axidilaton} $\Phi= a +i f_\varphi
\exp(-\varphi/f_\varphi)$, cf. Refs. \cite{Se92,MS01,Ka06}.
Moreover, the torsion-induced {\em quintaxion} may simulate `dark
energy' (DE) with interesting repercussions on the cosmological
evolution \cite{MR06}. The anharmonic oscillating phases of
inflationary expansion, decelerating graceful exit and acceleration
could be related to the observed epochs in our Universe.

In general, an  {\em induced} cosmological constant
\begin{equation}
\Lambda_\theta= -
{{3\theta_{\rm T}^* }\over{2\ell^2(
\theta_{\rm L}^\star +b_6)}}
         \label{coseff}
\end{equation}
of microscopic origin \cite{Mie84} is  unavoidable with an
interesting (Anti-) de Sitter background, resembling the intriguing
AdS/CFT correspondence. Moreover, there are strong indications from
supernova observations \cite{As05} that the present epoch of the
Universe is dominated by `dark energy'  in form of a tiny
cosmological constant. Moreover, a  coupling of the Euler or
Gauss-Bonnet term to a hypothetical scalar field may  generate
\cite{CE07} some of the cosmological dynamics responsible for the
transition from matter dominance to the  acceleration of the present
epoch associated with DE.

Consequently,  there is still a valid avenue to a consistent
quantization based on a topological version of selfdual SKY gravity,
departing, in a gauge covariant approach, from a $d$-exact
topological term. Due to the nilpotency of the corresponding BRST
charges \cite{MR03}, the $s$-exact term can easily account for the
necessary gauge constraints such as (\ref{dodual}) implying
Einsteinian gravity for the classical `background'. This, to some
extent,  provides an answer to the issue already raised 1963 by
Feynman \cite{FM95}, whether Einstein's GR, in view of  its
force-free geometrical concepts, needs to be quantized at all or if
curved spacetime can be left as an arena for quantized (topological)
fields to play out. One would like to see generalizations of our
approach to affine topological gauge models of gravity based on
superconnections \cite{Ne97,NS05} including the Higgs field.

\begin{theacknowledgments}
I would like to thank Al\'{\i} A. Rinc\'on Maggiolo, Eric S. Romero
and Dmitri Vassiliev for valuable comments. Moreover, (E.W.M.)
acknowledges the support of the SNI and thanks Noelia,  Markus
G\'erard Erik, and Miryam Sophie Naomi for encouragement.
\end{theacknowledgments}

\section{Appendix A: Riemann-Cartan geometry in Clifford
algebra-valued exterior forms}
The coframe $\vartheta^{\alpha} = e_{j}{}^{\alpha}\,dx^{j}$ of
dimension [length] and the dimensionless
 connection one-form $\Gamma_{\alpha}{}^{\beta}
 = \Gamma_{i\alpha}{}^{\beta}\,dx^{i}$
are the gauge potentials of nonlinearly realized {\em local
translations} \cite{TM00} and {\em local linear transformations},
respectively. The  dual basis
$\{\eta_{\alpha},\,\eta_{\alpha\beta},\,\eta_{\alpha\beta\gamma}
 \,,\eta_{\alpha\beta\gamma\delta}\}$
 of exterior forms can  be generated from the volume four-form
$\eta  = \eta_{\alpha\beta\gamma\delta}\,
\vartheta^{\alpha}\wedge\vartheta^{\beta}\wedge
\vartheta^{\gamma}\wedge\vartheta^{\delta}/4!$
 by consecutive interior products $\rfloor$ via
$\eta_{\alpha}:=  e_{\alpha}\rfloor\eta =\, ^{*}\vartheta_{\alpha}$
, etc. On a four--dimensional manifold with metric index ${\rm
sig}$, the Hodge dual of $p$--forms is almost involutive, i.e. $
^{**}\alpha = (-1)^{p(4-p)+{\rm sig}}\alpha$.  For spacetimes where
${\rm sig}=1$ holds, it induces an {\em almost complex structure},
cf. \cite{Br75}. In four dimensions, the {\em Hodge dual} applied to
two--forms is {\em conformally invariant} \cite{AKHS78}.

For a concise formulation of the BRST transformations it is
instrumental to reexpress the Riemann-Cartan structure of spacetime
in terms of Clifford algebra-valued differential forms: In the
familiar Pauli representation \cite{BD64}, the 16 matrices
$\{{{\mathbf 1}}_4,\gamma_\alpha,\sigma_{\alpha\beta},\gamma_5,
\gamma_5\gamma_\alpha\}$, where $\sigma_{\alpha\beta}:={i\over 2}
(\gamma_\alpha\gamma_\beta-\gamma_\beta\gamma_\alpha)$ are the
 Lorentz generators and
 $\gamma_5=-i\,\gamma_{\hat{0}}\,\gamma_{\hat{1}}\,\gamma_{\hat{2}}\,
 \gamma_{\hat{3}}$, constitute
 a basis of the {\em Clifford algebra}  in
four dimensions  with the defining relation
 \begin{equation}
\gamma_\alpha\gamma_\beta+\gamma_\beta\gamma_\alpha
  =2o_{\alpha\beta} \, {{\mathbf 1}}_4 \,.
\label{Cliff}
\end{equation}
They
are normalized by $Tr(\gamma_\alpha\; \gamma_\beta) =4\,o_{\alpha\beta}$   and
$Tr(\sigma_{\alpha\beta}\; \sigma^{\gamma\delta}) =
  8\,\delta_{[\alpha}^{\gamma}\; \delta_{\beta]}^{\delta}$,
where
$[\alpha\, \beta ]=\frac{1}{2}(\alpha\beta-\beta\alpha)$ denotes the
antisymmetrization of indices.

In terms of the Clifford algebra--valued coframe and  {\em connection}
\begin{equation}
\gamma:=\gamma_\alpha\vartheta^\alpha\,, \qquad
 \Gamma := {i\over 4} \Gamma^{\alpha\beta}\,\sigma_{\alpha\beta} =
 \Gamma^{\{\}} - K\, ,
\label{cofr}
\end{equation}
where $K:= {i\over 4} K^{\alpha\beta}\,\sigma_{\alpha\beta}$ is the
contortion one-form,  the $\overline{SO}_\circ(1,3)\cong
SL(2,C)$--covariant exterior derivative $D=d+[\Gamma,\quad]$ employs
the algebra--valued {\em form commutator} $[\Psi, \Phi] :=
\Psi\wedge\Phi - (-1)^{p_1p_2} \Phi\wedge\Psi$. The Hilbert-de
Donder and Lorentz type gauge conditions involve the Hodge dual and
can be rewritten as the following conditions on four-forms
\begin{equation}
d\, ^* \gamma= 0 \, , \qquad d\, ^* \Gamma= 0\, .
\end{equation}

Differentiation of these basic variables leads to the Clifford
algebra--valued {\em torsion} and {\em curvature} two--forms \ba
\Theta &:=&D\gamma =T^{\alpha}\gamma_{\alpha}= (d\vartheta^{\alpha}
+
\Gamma_{\beta}{}^{\alpha}\wedge\vartheta^{\beta})\gamma_{\alpha}\;,\nonumber\\
\Omega &:=& d\Gamma +\Gamma\wedge \Gamma = \frac{i}{
4}R^{\alpha\beta}\,\sigma_{\alpha\beta}=\frac{i}{ 4}(
d\Gamma_{\alpha}{}^{\beta} -
\Gamma_{\alpha}{}^{\gamma}\wedge\Gamma_{\gamma}{}^{\beta})\,
\sigma^{\alpha}{}_{\beta}\, , \label{torClif} \ea
 respectively. The compact Clifform formula for
the curvature was already known to Schr\"odinger \cite{Schro}. In a
RC spacetime, the translational and Lorentz--rotational {\em
Chern--Simons terms} read \ba C_{\mathrm{ TT}}  &:=&
{1\over{8\ell^2}} Tr\, ( {\gamma} \wedge {\Theta} )=
{1\over{2\ell^2}}\, {\vartheta^\alpha}\wedge
T_{\alpha}=-\frac{(-1)^{\rm sig}}{2 \ell^{2}} \,^*{\mathcal A}
\,,\nonumber\\
 C_{\mathrm{ RR}}  &:=& {\rm Tr}\, \left( {\Gamma}\wedge
{\Omega} - {1\over 3} {\Gamma}\wedge {\Gamma}\wedge {\Gamma} \right)
\, , \label{CRR} \ea where ${\cal A}:= \,^*(\vartheta_\alpha\wedge
T^\alpha)={\cal A}_i dx^i$ is the {\em axial torsion} one-form. The
translational Chern-Simons term is not Weyl invariant, cf. (3.14.9)
of Ref. \cite{HMMN95}, due to the occurence of a fundamental length
$\ell$. In view of (\ref{Cliff}), the Clifford algebra approach has
the advantage in employing the trace in the definition (\ref{CRR}),
whereas the usual generators $P_{\alpha}$ of translations commute
and do not have a non-degenerate Cartan--Killing metric.

The Ricci identity for a p-form  $\Psi$ reads \be D D \Psi =
[\Omega,\Psi] \label{ric} \,. \ee In RC geometry,  the first and
second {\em Bianchi identities} adopt the form
\begin{equation}
D\Theta \equiv [\Omega ,\gamma]\, ,\qquad\qquad D\Omega\equiv 0
,\qquad\qquad D\Omega^{(\star)}\equiv 0 \label{Bia}\, ,
\end{equation}
respectively, cf. Ref. \cite{Mie01} for further details. Here the
{\em Lie dual} of the curvature is defined by $
R_{\alpha\beta}^{(\star)}:= {1\over 2}\eta_{\alpha\beta\gamma\delta}
R^{\gamma\delta}$.

The Lagrangians  corresponding to the Bianchi identities (\ref{Bia})
are the boundary terms \ba L_{\rm NY}&:=&dC_{\rm TT} =
{1\over{2\ell^2}} \left(T^{\alpha}\wedge
T_\alpha+R_{\alpha\beta}\wedge\vartheta^\alpha
\wedge\vartheta^\beta\right)\, , \label{NiehYan}\\L_{\rm Pontr}&:=&
dC_{\rm RR}= {1\over 2}R_{\alpha}{}^{\beta}\wedge
R_{\beta}{}^{\alpha}\\
&=& -{1\over 2}R^{\{\}}_{\alpha\beta}\wedge R^{\{\}\alpha\beta} -
\frac{1}{12}d\left[\,^*{\mathcal A}\wedge R^{\{\}}
-\frac{1}{3}{\mathcal A}\wedge d{\mathcal A} +\frac{1}{9}
\,^*{\mathcal A}\wedge^*({\mathcal A} \wedge  \,^*{\mathcal
A})\right] . \nonumber\label{RPontr} \ea The latter contains,
amongst others, a term proportional to the Riemannian curvature
scalar $R^{\{\}}:=\,^*(R^{\{\}\alpha\beta}\wedge
\eta_{\beta\alpha})$ and the axial torsion piece $d{\cal A}\wedge
d{\cal A}$ of the axial anomaly \cite{KM01,Mi06} with a relative
factor 9. Up to normalizations, the four-forms (\ref{NiehYan}) and
(\ref{RPontr}) are  known as Nieh--Yan \cite{NY} and gravitational
Pontrjagin term, respectively.

On the other hand, the topological  Euler  term
\begin{eqnarray}L_{\rm Euler}&:=&(-1)^{{\rm sig}+1}
{\rm Tr}\{\Omega\wedge\Omega^{(\star)}\}=\frac{(-1)^{\rm
sig}}{2}\,R^{\alpha\beta}\wedge
 R_{\alpha\beta}^{(\star)}\nonumber\\
&=& \frac{(-1)^{\rm sig}}{2}\, d \big( \Gamma_{\alpha\beta}\wedge
R^{\alpha\beta(\star)} - {1\over 3}
\Gamma_\alpha{}^{\beta(\star)}\wedge \Gamma_\beta{}^\gamma\wedge
\Gamma_\gamma{}^\alpha\big)\nonumber\\
&\equiv& \frac{1}{2}\,R_{\alpha\beta}\wedge \,^* R^{\alpha\beta}
-2Ric_{\alpha\beta}\wedge \,^*Ric^{\alpha\beta} + \frac{1}{2}
Ric_{\alpha}{}^{\alpha}\wedge \,^*Ric_{\beta}{}^{\beta}
\label{Euler}
\end{eqnarray}
for Riemann-Cartan spaces has an equivalent representation in terms
of Yang's Lagrangian $L_{\rm SKY}$ as well as a Ricci-squared and a
curvature scalar squared term, cf. Eq. (3.1) of Ref. \cite{Mi81}.
The expression in terms of  the symmetric Ricci tensor, i.e.  the
zero-form $Ric_{\alpha\beta}:= (-1)^{\rm sig} \, ^*
(R_{(\alpha}{}^\delta \wedge \eta_{\delta\vert\beta)})$, is also
known as Gauss-Bonnet term.

Due to the algebraic {\em Lanczos identity} \cite{La38}, the {\em
double dual curvature} \be
  \,^*R_{\alpha\beta}^{(\star)}\equiv(-1)^{\rm sig}R_{\alpha\beta} +
  e_{[\alpha}\rfloor  G_{ \beta ]} + \frac{1}{4} R \eta_{\alpha\beta} +
  D_{[\alpha}  T_{\beta]}
 \label{DDual}
 \ee
 can be written  in terms of a  contraction of the EC three-form $G_\alpha:= R^{\beta\gamma}\wedge
\eta_{\alpha\beta\gamma}/2$ and the curvature scalar
 $R:= e_\beta\rfloor e_\alpha\rfloor R^{\alpha\beta}$, a zero form;
  cf. Ref. \cite{Mi81} for the same result in Riemannian spacetime and in components.

Likewise, the Einstein-Cartan three-form \ba
G &:=&\frac{1}{2}R^{\beta\gamma}\wedge\eta_{\alpha\beta\gamma}\gamma^\alpha \\
&=& G^{\{\}} + \frac{(-1)^{\rm sig}}{12}\left( e_\alpha\rfloor {\cal
A} \wedge\, ^*{\cal A}- \frac{1}{3}{\cal A} \wedge e_\alpha\rfloor\,
^*{\cal A}\right)\gamma^\alpha
 +\frac{(-1)^{\rm sig}}{6} \gamma\wedge d{\cal A}\nonumber\label{Eindec}
\ea
 decomposes into the Einstein three-form
 $G^{\{\}}=G_\alpha{}^\beta \eta_\beta\gamma^\alpha$ with respect to
 the Riemannian connection $\Gamma^{\{\}}$
and axial torsion pieces, see Ref. \cite{MR06} for details.

\section{Appendix B: BRST transformations}
 In the BRST formalism, the infinitesimal gauge transformations
 are converted, via ghosts, into operator transformations. Let
 $c:= {i\over 4} c^{\alpha\beta}\,\sigma_{\alpha\beta}$ denote
  the zero-form of the usual Faddeev-Popov ghost, cf.
  Refs. \cite{MR03,VH05},
  $\Psi:={i\over 4}\Psi^{\alpha\beta}_j\sigma_{\alpha\beta}dx^j$
  the topological ghost one-form  and
 $\Phi:={i\over 4}\Phi^{\alpha\beta}\sigma_{\alpha\beta}$ the
 corresponding ghost of the topological ghost. All are Lie
 algebra-valued due to the appearance of the generator
 $\sigma_{\alpha\beta}$ of the linear (or Lorentz) group.

 Then the global BRST transformations generated by the zero-form
 $s$ take the form
 \begin{eqnarray}
  s\Gamma &=& \Psi -Dc \,,\nonumber\\
  s c &=& \Phi -\frac{1}{2} [c, c] \,,\nonumber\\
  s\Omega &=& D\Psi -[c,\Omega] \,,\nonumber\\
  s\Psi &=& -D\Phi -[c,\Psi] \,, \nonumber\\
   s\Phi&=&  -[c,\Phi] \,. \label{BRSTtr}
  \end{eqnarray}
  This is consistent with the interpretation of $\Gamma$ and $\Omega$
  as connection one-form and  gauge two-form, respectively.
  The topological ghost $\Psi$ complements the
  inhomogeneous  transformation law of infinitesimal gauge fields. By construction,
  these BRST transformations are nilpotent for all
  variables, i.e. $s^2=0$.

In the rather elegant geometrical interpretation of Ref. \cite{BS88},
  the {\em graded}\footnote{The grading permits using the direct
  sum $\oplus$ of exterior forms carrying
  different form degree $p$ and ghost number $g$, such that the graded
  commutator is now defined by
  $[\Psi, \Phi] :=
\Psi\wedge\Phi - (-1)^{(p_1+g_1)(p_2+g_2)} \Phi\wedge\Psi$, cf. Ref.
\cite{CS92}.}
 connection and curvature  defined via
 \begin{equation}
  \widetilde\Gamma:=\Gamma \oplus c \,,\qquad   \widetilde\Omega:=\Omega\oplus
   \Psi \oplus \Phi\,,\label{gradCartan}
   \end{equation}
 satisfy the corresponding {\em graded} structure equation and
 second  Bianchi identity for the graded curvature, i.e.
  \begin{eqnarray}
  (d \oplus s)\widetilde\Gamma +
  \frac{1}{2}[\widetilde\Gamma,\widetilde\Gamma]&=&\widetilde\Omega\,, \nonumber\\
  (d \oplus s)\widetilde\Omega +[\widetilde\Gamma,\widetilde\Omega] &\equiv& 0
  \,.
  \label{gradBianchi}
 \end{eqnarray}
  They comprise all the BRST transformations (\ref{BRSTtr}) above,
  and constitute an ordinary de Rham cohomology. Moreover,
 a straightforward proof of the nilpotency of the BRST operator $s$ now
 follows simply from $(d \oplus s)(d \oplus s)\equiv 0$ as a result of the graded
Bianchi
 identity (\ref{gradBianchi}), the anti-commutation of the graded
 commutator $[s,d]=0$,  and the Poincar\'e lemma $dd\equiv0$.

In order to implement the gauge constraints one uses the antifield
formalism, where the Lorentz algebra-valued antighosts $
\overline{c}$, $ \overline{\chi}$, and $ \overline{\Phi}$  obey the
following BRST transformation rules
\begin{eqnarray}
  s \overline{c} &=& B\,, \qquad s B=0\,,\nonumber\\
  s \overline{\chi} &=& b\,,\qquad s b=0\,, \nonumber\\
  s \overline{\Phi} &=& \overline{\eta}\,,    \qquad s \overline{\eta}=0 \,.
  \end{eqnarray}
By construction $s$ is nilpotent for the antighosts, since the
Lagrangian multiplier $\overline{\eta}$, the selfdual two-forms $B$
and $b$ are  auxiliar fields introduced as trivial pairs.  A
symmetric ghost/antighosts spectrum of an extended BRST invariance
could be obtained via a field redefinition \cite{BG00}. (For
metric-affine gravity, an antifield formalism has been developed in
Ref. \cite{Gr98} without employing topological ghosts.)

By introducing a BRST gauge field  $\alpha= \alpha_i dx^i$ with
ghost number $-1$ and a commuting ghost $\lambda$ of $\alpha$, one
can promote \cite{CB92} the global  BRST transformations
(\ref{BRSTtr}) into {\em local} ones, where
\begin{equation}
s_{\rm l}(\alpha+\lambda) =- d\lambda \,, \qquad s_{\rm l}\lambda =0
\end{equation}
is  satisfying the algebra
\begin{equation}
(d \oplus s_{\rm l})(\alpha+\lambda) =d\alpha.
\end{equation}
The cohomology (\ref{gradBianchi}) of the BRST transformation
remains unchanged by this promotion which, likewise, can be
generated via  the field redefinitions $c\rightarrow
(\alpha+\lambda)c$, $\Psi\rightarrow (\alpha+\lambda)\Psi$ as well
as $\Phi\rightarrow (\alpha+\lambda)^2\Phi$ of the ghosts, see Ref.
\cite{CB92} for more details. Thus, local BRST invariance of an
action puts no more restrictions on its form than the usual global
one:  The gauge field $\alpha$ is only present to compensate for the
enlargement of the symmetry, from global to local, but it cannot
propagate, due to its non-vanishing ghost number.

\section{Appendix C: BRST translations and diffeomorphisms}
By gauging the translational part $R^4$ of the affine group
\cite{HMMN95}, there arises a translational connection $\Gamma^{(\rm
T)}$ which in gravity is usually `soldered' \cite{TM00} to the base
manifold, with the familiar tetrads or coframe $\gamma$ as the
result, cf. (\ref{cofr}). Then the topological structure equations
(\ref{gradBianchi}) for the linear connection get amended by the
corresponding {\em graded} first Cartan structure equation and the
 first Bianchi identity
  \begin{eqnarray}
  (d \oplus s)\gamma +
  [\widetilde\Gamma,\gamma]&=&\widetilde\Theta\,, \nonumber\\
  (d \oplus s)\widetilde\Theta +[\widetilde\Gamma,\widetilde\Theta]
  &\equiv& [\widetilde\Omega,\gamma]\, ,
   \label{grad1Bianchi}
 \end{eqnarray}
respectively, for the graded torsion two-form $
\widetilde\Theta:=\Theta \,\oplus\,
   \psi \,\oplus \,\phi$.
 (When undoing the `soldering', the translational connection $\Gamma^{(\rm T)}$
 would need to be graded as well, e.g. by the substitution  $\gamma \rightarrow \widetilde\gamma:=\gamma
\oplus c$ in the first structure equation.)

 In topological models of gravity, diffeomorphisms can also be
taken account of by generalizing the BRST transformations $s$ via
$s\rightarrow \widetilde s = s+ \L_\xi$ involving the covariant Lie
derivative $\L_\xi:= \xi\rfloor D - D\xi\rfloor$ built from the
interior product $\rfloor$ with respect to an anticommuting ghost
vector field $\xi=\xi^i\partial_i$. Note that there  is sign
difference in the definition of $\L_\xi$, since $\xi$ has ghost
number one. Then the BRST algebra remains intact \cite{BT02,CS92},
up to a redefinition of all ghosts by means of a similarity
transformation generated by the formal exponential\be
\exp(\xi\rfloor) := {\mathbf 1} +\xi\rfloor +\frac{1}{2!}
\xi\rfloor\xi\rfloor +\frac{1}{3!} \xi\rfloor\xi\rfloor\xi\rfloor
+\cdots .\ee In effect, the graded
 curvature and torsion in the cohomologies (\ref{gradBianchi}) and
(\ref{grad1Bianchi}) are replaced by $\exp(\xi\rfloor)\widetilde
\Omega$ and $\exp(\xi\rfloor)\widetilde \Theta$, respectively.



\begin{thebibliography}{999}
\vskip 30pt
\bibitem{AB01}
  A.~D.~Alexeev, K.~A.~Bronnikov, N.~I.~Kolosnitsyn, M.~Y.~Konstantinov, V.~N.~Melnikov and A.~J.~Sanders,
  ``Measurement of the gravitational constant G in space (Project SEE):
  Sensitivity to orbital parameters and space charge effect,''
  arXiv:gr-qc/0104066.
\bibitem{CB92}
C.~Aragao de Carvalho and L.~Baulieu:
  ``Local BRST symmetry and superfield formulation of the Donaldson-Witten
  theory,'' Phys.\ Lett.\ B {\bf 275}, 323 (1992).
\bibitem{As05}
P.~Astier et al.,
  ``The Supernova Legacy Survey: Measurement of $\Omega_{\rm M}$,
  $\Omega_\Lambda$ and $w$ from the first year data set,''
   Astronomy and Astrophysics {\bf 447}, 31 (2006).

\bibitem{AKHS78}M.F. Atiyah, N.J. Hitchin, and I.M. Singer, Proc. R. Soc.
(London) {\bf A 362}, 425 (1978).

\bibitem{Ba85}
  L.~Baulieu: ``Perturbative gauge theories,''
  Phys.\ Rept.\  {\bf 129}, 1 (1985).

\bibitem{BS88}
  L.~Baulieu and I.~M.~Singer:
  ``Topological Yang-Mills symmetry,''
  Nucl.\ Phys.\ Proc.\ Suppl.\  {\bf 5B}, 12 (1988).
\bibitem{BT02}
  L.~Baulieu and A.~Tanzini:
  ``Topological gravity versus supergravity on manifolds with special holonomy,''
  JHEP {\bf 0203}, 015 (2002).
 \bibitem{BD81}
  I.~M.~Benn, T.~Dereli and R.~W.~Tucker,
  ``Double dual solutions of generalized theories of gravitation,''
  Gen.\ Rel.\ Grav.\  {\bf 13}, 581 (1981).
\bibitem{BB91}
  D.~Birmingham, M.~Blau, M.~Rakowski and G.~Thompson:
  ``Topological field theory,'' Phys.\ Rept.\  {\bf 209}, 129 (1991).

\bibitem{BD64}J.D.
Bjorken  and S.D. Drell: {\em Relativistic Quantum Mechanics}
(Mc Graw--Hill, New York 1964).
\bibitem{BT90}
  M.~Blau and G.~Thompson:
  ``Do metric independent classical actions lead to topological field
  theories?,'' Phys.\ Lett.\ B {\bf 255}, 535 (1991).

\bibitem{BG00}
  N.~R.~F.~Braga and C.~F.~L.~Godinho:
  ``Extended BRST invariance in topological Yang-Mills theory revisited,''
  Phys.\ Rev.\ D {\bf 61}, 125019 (2000).

\bibitem{Br75}C.H. Brans,  J. Math. Phys. {\bf 15}, 1559 (1974);
{\bf 16}, 1008 (1975).

\bibitem{CS92}
  L.~N.~Chang and C.~P.~Soo:
  ``BRST cohomology and invariants of four-dimensional gravity in Ashtekar variables,''
  Phys.\ Rev.\ D {\bf 46}, 4257 (1992).
\bibitem{CE07}
  G.~Cognola, E.~Elizalde, S.~Nojiri, S.~Odintsov and S.~Zerbini,
  ``String-inspired Gauss-Bonnet gravity reconstructed from the universe
  expansion history and yielding the transition from matter dominance to dark
  energy,''
  Phys.\ Rev.\  D {\bf 75}, 086002 (2007).

\bibitem{DT02}
T.~Dereli and R.~W.~Tucker:
``A broken gauge approach to gravitational mass and charge,''
JHEP {\bf 0203}, 041 (2002).

 \bibitem{EGH80}
T.~Eguchi, P.~B.~Gilkey and A.~J.~Hanson:
``Gravitation, gauge theories and differential geometry,''
Phys. Rept. {\bf 66}, 213 (1980).

 \bibitem{Fa96}L.D. Faddeev: ``How we understand ``quantization" a
 hundred years after Max Planck", Phys. Bl. {\bf 52}, 689 (1996).

 \bibitem{Fa76} E.~E.~Fairchild: ``Gauge theory of gravitation,''
  Phys.\ Rev.\  D {\bf 14},  384 (1976).
\bibitem{Fa77}
  E.~E.~Fairchild:
  ``Yang-Mills formulation of gravitational dynamics,''
  Phys.\ Rev.\  D {\bf 16},  2438 (1977).
  \bibitem{FM95}
  R.~P.~Feynman, F.~B.~Morinigo, W.~G.~Wagner and B.~Hatfield (eds):
  ``Feynman lectures on gravitation,''
(Addison-Wesley, Reading, USA 1995) 232 p.
  \bibitem{GH78}
C.~H.~Gu, H.~S.~Hu, D.~Q.~Li, C.~L.~Shen, Y.~L.~Xin and C.~N.~Yang:
``Riemannian spaces with local duality and gravitational
instantons,'' Sci.\ Sin.\  {\bf 21}, 475 (1978).
 \bibitem{Gr98}
  F.~Gronwald:
  ``BRST-antifield treatment of metric-affine gravity,''
  Phys.\ Rev.\ D {\bf 57}, 961 (1998).

\bibitem{GN98}
B.~S. Guilfoyle and B.~C. Nolan:
``Yang's gravitational theory,'' Gen. Rel. Grav.  {\bf 30}, 473 (1998).


\bibitem{HMMN89}
F.W. Hehl,  J.D. McCrea, E.W. Mielke, and Y. Ne'eman:
``Progress in metric--affine gauge theories of gravity with local scale
        invariance", Found. Phys. {\bf 19}, 1075 -- 1100 (1989).


\bibitem{HMMN95} F.W. Hehl, J.D. McCrea, E.W. Mielke, and Y. Ne'eman,
Phys. Rept. {\bf 258},  1- 171 (1995).

\bibitem{Hi59}P.W. Higgs, Nuovo Cimento {\bf 11}, 816 (1959).
\bibitem{KC07}
  D.~J.~Kapner, T.~S.~Cook, E.~G.~Adelberger, J.~H.~Gundlach, B.~R.~Heckel, C.~D.~Hoyle and H.~E.~Swanson,
  ``Tests of the gravitational inverse-square law below the
  dark-energy length
  scale,''
  Phys.\ Rev.\ Lett.\  {\bf 98}, 021101 (2007).
\bibitem{Ka06}
  R.~K.~Kaul:
  ``Gauge theory of gravity and supergravity,''
  Phys.\ Rev.\  D {\bf 73}, 065027 (2006).

\bibitem{KS86}
T.W.B. Kibble and K.S. Stelle: ``Gauge theories of gravity and
supergravity", in {\em  Progress in Quantum Field Theory},
Festschrift for Umezawa, H. Ezawa and S. Kamefuchi, eds. (Elsevier
Science Publ. Amsterdam 1986), p. 57.

\bibitem{KN61}C.W. Kilmister and D.L. Newman, Proc.
Cambridge Phil. Soc. (Math. Phys. Sci.) {\bf 57},  851 (1961).
 \bibitem{KM01}
 D. Kreimer and E.W. Mielke: ``Comment on: Topological
invariants, instantons, and the chiral  anomaly on spaces with
torsion", Phys. Rev. {\bf D63}, 048501 (2001).
\bibitem{Ku86} R. Kuhfu{\ss} and J. Nitsch,   Gen. Rel. Grav. {\bf 18},
1207 (1986).
\bibitem{LP88}
  J.~M.~F.~Labastida and M.~Pernici,
  ``A Lagrangian for topological gravity and its  BRST quantization,''
  Phys.\ Lett.\  B {\bf 213}, 319 (1988).
\bibitem{La38}C.  Lanczos:
``A remarkable property of the Riemann--Christoffel
tensor in four dimensions",  Ann. Math. {\bf 39},  842 (1938).
 \bibitem{LN90}
C.~Y.~Lee and Y.~Ne'eman: ``Renormalization of gauge affine
gravity,'' Phys. Lett. B {\bf 242}, 59 (1990).

\bibitem{MM77}
S.~W.~MacDowell and F.~Mansouri:
``Unified geometric theory of gravity and supergravity,''
Phys. Rev. Lett.   {\bf 38}, 739 (1977) [Erratum-ibid.\  {\bf 38}, 1376 (1977)].

\bibitem{Mi81}E.W. Mielke:
``On pseudoparticle solutions in Yang's theory of gravity", Gen.
Rel. Grav. {\bf 13},  175 -- 187 (1981).
 \bibitem{Mie84} E. W. Mielke,  J. Math. Phys. {\bf 25}, 663 (1984).
\bibitem{Mi84b} E. W. Mielke:  ``On pseudoparticle solutions in the
Poincar\'e gauge theory of gravity", Fortschr. Phys. {\bf 32}, 639
(1984).

\bibitem{Mie91}
 E.W. Mielke:
 ``Ashtekar's complex variables in general relativity and its teleparallelism
 equivalent", Ann. Phys. (N.Y.) {\bf 219}, 78-- 108 (1992).
\bibitem{Mie01}E. W.
Mielke: ``Beautiful gauge field equations in Clifforms", Int. J.
Theor. Phys. {\bf 40}, 171 -- 190  (2001).

\bibitem{Mi06}
E.W. Mielke: ``Duality and renormalization scheme for
Einsteinian gravity as a fix point within a gravitational gauge
framework", Electronic Journal of Theoretical Physics (EJTP) {\bf
3}, No. 12, 1-18 (2006).
\bibitem{MiA06}
E.W. Mielke: ``Anomalies and gravity", in: {\em Particles and
Fields}, Commemorative Volume of the Division of Particles and
Fields of the Mexican Phys. Soc., Morelia Michoac\'an, 6-12 Nov.
2005, Part B., M.A. P\'erez, L.F. Urrutia, and L. Villase\~nor, eds.
(AIP Conference Proc., Melville N.Y. 2006) Vol. 857, pp. 246 -- 257.

\bibitem{MR03}
E.W.  Mielke and  A. A. Rinc\'on Maggiolo:
``Algebra for a BRST quantization of metric-affine gravity",
 Gen. Rel. Grav. {\bf 35}, 771-789 (2003).
 \bibitem{MR05}
 E.W.  Mielke and  A. A. Rinc\'on Maggiolo:
``Duality in Yang's theory of gravity",
Gen. Rel. Grav. {\bf 37}, 997-1007 (2005).
\bibitem{MS01}
  E.~W.~Mielke and F.~E.~Schunck,
  ``Are axidilaton stars massive compact halo objects?,''
  Gen.\ Rel.\ Grav.\  {\bf 33}, 805 (2001).
 \bibitem{MR06}  E.~W. Mielke and E.~S.~Romero:
  ``Cosmological evolution of a torsion-induced quintaxion,''
  Phys. Rev. D {\bf 73}, 043521 (2006).

 \bibitem{NSO91}
A. Nakamichi, A. Sugamoto, and I. Oda, Phys. Rev. {\bf D 44}, 3835
(1991).
\bibitem{Ne97}
Y.~Ne'eman: ``A superconnection for Riemannian gravity as spontaneously
broken  SL(4,R) gauge theory,''
Phys.\ Lett.\ B {\bf 427}, 19 (1998).
\bibitem{NS05}
  Y.~Ne'eman, S.~Sternberg and D.~Fairlie,
  ``Superconnections for electroweak su(2/1) and extensions, and the mass of
  the Higgs,''
  Phys.\ Rept.\  {\bf 406}, 303 (2005).

 \bibitem{NY} H.T. Nieh and M.L. Yan,
J. Math. Phys. {\bf 23},  373--374 (1982).

\bibitem{OH96}
  Y.~N.~Obukhov and F.~W.~Hehl,
  ``On the relation between quadratic and linear curvature Lagrangians in
  Poincar\'e gauge gravity,''
  Acta Phys.\ Polon.\  B {\bf 27}, 2685 (1996).

\bibitem{Pa83}
H.~R.~Pagels: ``Gravitational gauge fields and the cosmological constant,''
Phys. Rev. D {\bf 29}, 1690 (1984).
\bibitem{PT93}
  M.~J.~Perry and E.~Teo,
  ``Topological conformal gravity in four-dimensions,''
  Nucl.\ Phys.\  B {\bf 401}, 206 (1993).
\bibitem{Schro} E.
Schr\"odinger: ``Diracsches Elektron im Schwerefeld I.",
Sitzungsber.   Preuss.  Akad. Wiss.  Phys.
Math. Kl. {\bf 11},  105 (1932).
\bibitem{Se92}
A. Sen, Phys. Lett. A {\bf 8},  2023 (1993).
\bibitem{Peter} E. Sezgin and P. van Nieuwenhuizen, Phys. Rev. {\bf
    D21},  3269--3280 (1980).
\bibitem{St77}K.S. Stelle, Phys. Rev. {\bf D16},  953 (1977).
\bibitem{St58}G. Stephenson, Nuovo Cimento {\bf 9},   263 (1958).
\bibitem{Th75} A.H. Thompson,  Phys. Rev. Lett. {\bf 34}, 505; {\bf 35}, 320
(1975).
\bibitem{TM00}
R. Tresguerres  and  E.W.~Mielke: ``Gravitational Goldstone fields
from affine gauge theory",  Phys. Rev. {\bf D62},  44004 (2000).

\bibitem{VH05}
  J.~W.~Van Holten,
  ``Aspects of BRST quantization,''
  Lect.\ Notes Phys.\  {\bf 659}, 99 (2005).


\bibitem{Va02}D. Vassiliev: ``Pseudoinstantons
in metric-affine field theory",
Gen. Rel. Grav. {\bf 34}, 1239 (2002).


\bibitem{We19}H. Weyl:
``Eine neue Erweiterung der Relativit\"atstheorie",
 Annalen Phys. (Leipzig) IV.  Folge, {\bf 59}, 103 (1919).

\bibitem{We29} H. Weyl: ``Gravitation and the electron",
 Proc. Nat. Acad. Sci. (Washington) {\bf 15}, 323 (1929).
\bibitem{Wi05}
  C.~M.~Will,
  ``Was Einstein right? Testing relativity at the centenary,''
  Annalen Phys.\  {\bf 15}, 19 (2005).
 \bibitem{Wi88}
  E.~Witten:``Topological quantum field theory,''
  Commun.\ Math.\ Phys.\  {\bf 117}, 353 (1988); {\bf 118}, 411 (1988).
\bibitem{WiG88}
  E.~Witten:``Topological gravity,''
  Phys.\ Lett.\  B {\bf 206}, 601 (1988).

 \bibitem{Ya74}
C. N. Yang: ``Integral formalism for gauge fields",
Phys. Rev. Lett. {\bf 33}, 445--447 (1974).


\bibitem{Zhytnikov} V.V. Zhytnikov, J. Math. Phys. {\bf 35},  6001--6017 (1994).
\end{thebibliography}
\end{document}